# Electric-field-induced formation and annihilation of skyrmions in two-dimensional magnet

Jingman Pang[1,2], Hongjia Wang[3], Yufei Tang[2*], Yun Zhang[3*], Laurent Bellaiche[4*]

[1]*Faculty of Chemistry and Chemical Engineering, Engineering Research Center of Advanced Ferroelectric Functional Materials, Baoji Key Laboratory of Advanced Functional Materials, Baoji University of Arts and Sciences, 1 Hi-Tech Avenue, Baoji, Shaanxi, P. R. China*

[2]*Shaanxi Province Key Laboratory of Corrosion and Protection, Department of Materials Science and Engineering, Xi'an University of Technology, Xi'an, Shaanxi, 710048, PR China*

[3]*College of Physics and Optoelectronic Technology, Collaborative Innovation Center of Rare-Earth Functional Materials and Devices Development, Baoji University of Arts and Sciences, Baoji 721016, China*

[4]*Physics Department and Institute for Nanoscience and Engineering, University of Arkansas, Fayetteville, Arkansas 72701, USA*

[*]E-mail: zhangyun_xtu@163.com  laurent@uark.edu

**Abstract:** Electric manipulation of skyrmions in 2D magnetic materials has garnered significant attention due to the potential in energy-efficient spintronic devices. In this work, using first-principles calculations and Monte Carlo simulations, we report the electric-field-tunable magnetic skyrmions in $MnIn_2Te_4$ monolayer. By adjusting the magnetic parameters, including the Heisenberg exchange interaction, DMI, and MAE, through applying an electric field, the formation or annihilation of skyrmions can be achieved. Our work suggests a platform for experimental realization of the electric-field-tunable magnetic skyrmions in 2D magnets.

# I. Introduction

Magnetic skyrmions are spin structures that exhibit a localized vortex-like configuration [1-5]. These structures are defined by the spins wrapping around a unit sphere and possessing an integer topological charge, endowing them with exceptional stability against transitions into trivial spin configurations due to their topological properties. As a result, magnetic skyrmions have great potential for applications in spintronics and next-generation information storage. The Dzyaloshinskii-Moriya interaction (DMI), which is an antisymmetric exchange interaction arising from spin-orbit coupling (SOC), plays an essential role in the formation of skyrmions [6, 7]. The existence of DMI is limited to systems that lack local inversion symmetry, including bulk magnets with chiral crystal structures [8-11] and multilayer films composed of magnetic and heavy-metal layers [12-16]. The latter have garnered significant attention as potential hosts for skyrmions due to their inherent inversion symmetry breaking and the ability to tune magnetic parameters by altering layer thickness and interface composition.

Recently, the discovery of long-range magnetic order in two-dimensional (2D) materials, represented by $CrI_3$, $CrGeTe_3$, and $Fe_3GeTe_2$, open up a new platform for exploring exotic magnetic phenomena down to the 2D scale [17-19]. Notably, the observation of skyrmions in 2D magnets $Cr_2Ge_2Te_6$ [20] and $Fe_3GeTe_2$ [21] was reported by two distinct experimental groups. However, the presences of skyrmions in thick layers of $Cr_2Ge_2Te_6$ and $Fe_3GeTe_2$ goes against expectations due to the point group of their monolayers, which would typically inhibit the occurrence of DMI [22]. **Rather than the DMI, it is found that the multiple fourth-order interactions are responsible for stabilizing the topological spin textures in $Fe_3GeTe_2$ [23].** Subsequently, several strategies in both theory and experiments have been proposed to break the symmetry of 2D magnetic materials, which can induce a significant DMI and facilitate the formation of skyrmions. In particular, stable skyrmion lattices are observed in 2D magnets van der Waals (vdW) heterostructures such as $Fe_3GeTe_2/WTe_2$, $Fe_3GeTe_2/Co/Pd$ multilayer, and $Fe_3GeTe_2/Cr_2Ge_2Te_6$, which can be attributed to the DMI induced by the interfacial proximity effect [24-28]. Another focus is the family of

Janus vdW magnets - characterized by the absence of inversion symmetry, with different atoms occupying the upper and lower layers - has been predicted to possess a DMI of significant magnitude, making them capable of generate skyrmions [29-33].

In order for skyrmions to be effectively utilized in practical applications, it is imperative not only to develop new materials that can host these magnetic structures, but also to explore and employ methods for manipulating and controlling them. The formation of magnetic skyrmions usually involves a delicate interplay between Heisenberg exchange, magnetic anisotropy (MAE) and DMI [34, 35]. Recent advances in controlling DMI have been made through the exploration of various stackings of heavy metal /ferromagnetic metal heterostructures and the introduction of ionic species such as oxygen and hydrogen via chemisorption [15, 16, 36-39]. It is noteworthy that, as a promising energy-efficient approach in low-power spintronics, the utilization of an electric field has been shown to enable the tuning of both the strength and chirality of DMI in multilayer films [40-43]. Moreover, by constructing 2D ferromagnetic/ferroelectric vdW heterostructures such as $Cr_2Ge_2Te_6/In_2Se_3$ [44] and $Fe_3GeTe_2/In_2Se_3$ [45], the DMI of 2D magnets can be controlled through ferroelectric polarization, similar to the effect of electric field. The induced interfacial DMI is rather insignificant in weak vdW heterostructures. **For instance, the |D/J| (D is DMI and J is Heisenberg exchange coupling) ratio is 0.04 for $Cr_2Ge_2Te_6/In_2Se_3$ and 0.075 for $Fe_3GeTe_2/In_2Se_3$. The |D/J| values in these systems are below the typical range of 0.1-0.2, which is considered to be the typical range in which skyrmionic phases are likely to be produced as either the ground state or metastate.** In addition, the large MAE of $Cr_2Ge_2Te_6$ and $Fe_3GeTe_2$ tends to force spins to align parallelly, resulting in the formation of a ferromagnetic ground state. Thus, an alternative approach to achieve electric field control of skyrmions in 2D magnets is to manipulate their MAE through the application of electric field. Experimental studies have already demonstrated that skyrmions can be created or annihilated in multilayer systems by controlling the MAE using a negative/positive voltage pulse [46]. Additionally, for 2D magnets, theoretical studies have shown that the MAE of the metallic ferromagnet $Fe_3GeTe_2$ can be dramatically changed by applying an external electric field, even allowing for the reorientation of the easy axis [47, 48]. Therefore, achieving electric-field tunable skyrmions in 2D magnetic materials is highly desired and could significantly advance the field of spintronics and data storage [49].

In the present work, via first-principles calculations and Monte Carlo simulations, we unveil that the dn-MnIn$_2$Te$_4$ monolayer (ML) is promising 2D magnet with skyrmionics physics. Despite having inherent large DMI, the pristine dn-MnIn$_2$Te$_4$ ML exhibits a large MAE that prevents it from holding a skyrmion state. However, applying an electric field results in a significant decrease in MAE and a gradual increase in the isotropic Heisenberg exchange couplings $J_1$ and $J_2$, whereas the DMI shows a negligible response. As a result, electrically controllable skyrmions can be created or annihilated by the applied electric field, leading to exotic skyrmionic phenomena. Such phenomenon is revealed to be closely related to the physical quantities of κ and $J_2/J_1$ ratio, which can describe the required conditions for such physics. Our work illustrates the potential of manipulating magnetic skyrmions in 2D magnets nonvolatility through electric field manipulation, which may have promising device applications.

**In the present work, we utilize first-principles calculations and Monte Carlo simulations to investigate the potential of dn-MnIn$_2$Te$_4$ monolayer (ML) as a promising 2D magnet with skyrmionics physics. Despite having an inherent large DMI, the pristine dn-MnIn$_2$Te$_4$ ML is found to have a MAE that prevents it from holding a skyrmion state. However, when an electric field is applied, the MAE is significantly reduced, and the isotropic Heisenberg exchange couplings $J_1$ and $J_2$ are gradually increased, while the DMI shows negligible response. As a result, skyrmions could be created or annihilated by the applied electric field, resulting in exotic skyrmionic phenomena. We further reveal that such phenomenon is closely related to the physical quantities of κ and $J_2/J_1$ ratio, which can describe the necessary conditions for such physics. Our work demonstrates the potential of using electric field manipulation to non-volatilely control magnetic skyrmions in 2D magnets, which holds promise for novel device applications.**

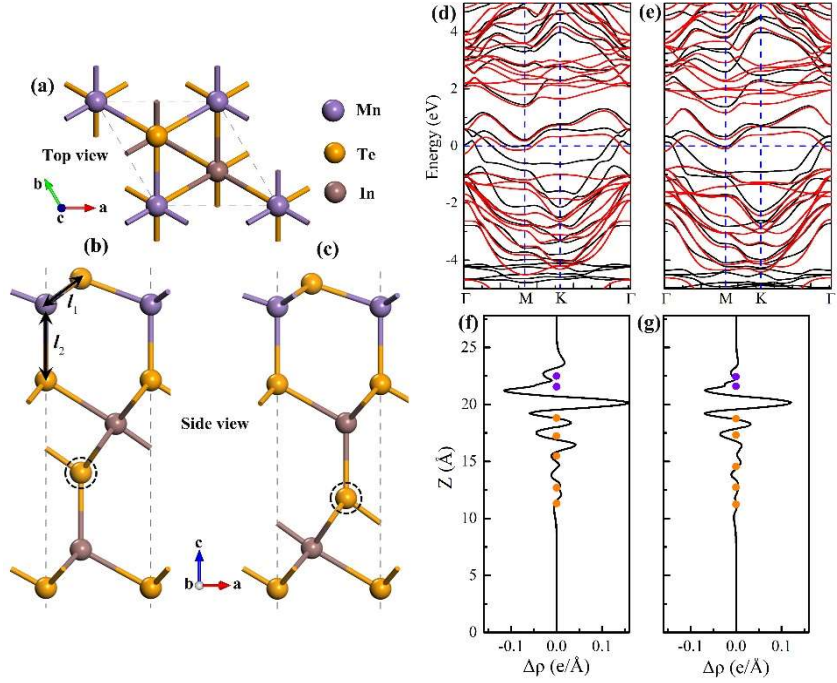

Figure 1. (a) Top view of the crystal structure of MnIn$_2$Te$_4$ MLs. Side views of (b) dn- and (c) up-MnIn$_2$Te$_4$ ML. Band structures of (d) dn- and (e) up-MnIn$_2$Te$_4$ ML. The black and red lines in (d, e) correspond to spin-up and spin-down states, respectively. The Fermi level is set to 0 eV. Planar-average charge-density difference of (f) dn- and (g) up-MnIn$_2$Te$_4$ ML. The purple and orange dots represent the positions of MnTe and In$_2$Se$_3$ atoms, respectively.

Figure 1 presents the crystal structure of MnIn$_2$Te$_4$ ML, which is derived from the prototype MIn$_2$S$_4$ (M =Zn, Mn or Fe) [50, 51]. The lattice constant of MnIn$_2$Te$_4$ ML is optimized to be 4.35 Å. The unit cell (UC) consists of one Mn atom, two In atoms, and four Te atoms, stacked in a sequence of Te-Mn-Te-In-Te-In-Te. The structure of MnIn$_2$Te$_4$ ML can also be viewed as a heterostructure, which is comprised of a ML In$_2$Te$_3$ on top of which 1 UC MnTe is placed. Therefore, the MnIn$_2$Te$_4$ ML, with the space group P3m1, obviously presents inversion asymmetry. Given the predicted ferroelectric nature of In$_2$Te$_3$ [52], the structure shown in Figure 1(b) can be regarded as having a downward polarization of In$_2$Te$_3$. This structure is denoted as dn-MnIn$_2$Te$_4$. Additionally, by shifting the Te atoms of the fifth layer downwards, we could construct a structure with an upward polarization of In$_2$Te$_3$, which is named as up-MnIn$_2$Te$_4$ (Figure 1(c)). **In order to assess the stability of these two distinct configurations of MnIn$_2$Te$_4$ ML**, the phonon spectra are first computed. As depicted in Figure S1 (a) and (b), we can only observe a tiny negative frequency in the vicinity of the Γ point, indicting their dynamic stabilities. We also calculate the independent elastic constants for dn-MnIn$_2$Te$_4$ and up-MnIn$_2$Te$_4$ ML, which are determined to be $C_{11}$ = 55.4 N m$^{-1}$ and $C_{12}$ = 29.4 N m$^{-1}$, and $C_{11}$ = 62.5 N m$^{-1}$ and $C_{12}$ = 32.8 N m$^{-1}$, respectively. All calculated values satisfy the Born−Huang criteria ($C_{11}$> 0, $C_{66}$> 0 (2$C_{66}$ = $C_{11}$ - $C_{12}$)), suggesting that these materials are mechanically stable [53]. Thus, it is expected that MnIn$_2$Te$_4$ will be synthesized experimentally using a method similar to that used for its prototype compounds ZnIn$_2$S$_4$ and FeIn$_2$S$_4$. In addition, as both structures possess broken inversion symmetry, it is anticipated that they would exhibit intriguing magnetic properties, such like skyrmions.

To explore the salient features of magnetism in our systems, we adopt the following spin Hamiltonian:

$$H = -\sum_{<i,j>} J_{ij}\mathbf{S}_i \cdot \mathbf{S}_j - \sum_{<i,j>} \mathbf{d}_{ij} \cdot (\mathbf{S}_i \times \mathbf{S}_j) - \sum_{<i,j>} \lambda(S_i^z S_j^z) \quad (1)$$
$$- \sum_i K(S_i^z)^2 - B\sum_i S_i^z$$

where $\mathbf{S}_i$ is the unit vector of a Mn atom at site i. The isotropic Heisenberg exchange coupling, denoted by $J_{ij}$, spans over all first and second nearest-neighbor (NN) Mn pairs. Second term $\mathbf{d}_{ij}$ represents the vector that characterizes the DMI for the first NN Mn pairs. The third term is the anisotropic symmetric exchange that runs over all the NN Mn sites. K is the single ion anisotropy (SIA). In the last term, $B$ indicates the strength of the external magnetic field along z axis.

Table I. The calculated magnetic parameters of dn- and up-MnIn$_2$Te$_4$ ML. $J$, d, λ, and K represent the isotropic Heisenberg exchange, DMI, anisotropic symmetric exchange, and single ion anisotropy, respectively. m is the magnetic moment of Mn atoms.

|  | $J_1$ (meV) | $J_2$ (meV) | $d_1$ (meV) | $|d_1/J_1|$ | λ (meV) | K (meV) | m ($\mu_B$) |
|---|---|---|---|---|---|---|---|
| dn-MnIn$_2$Te$_4$ | 3.184 | -0.456 | 0.726 | 0.228 | 0.217 | -0.048 | 4.41 |
| up-MnIn$_2$Te$_4$ | -6.312 | 0.184 | 0.224 | 0.035 | 0.301 | 0.893 | 4.37 |

The coefficients of Eq. (1) are listed in Table I and are extracted from DFT calculations and the energy mapping method (see Note 1 in the Supporting Information for details). As one can see from Table I, the NN Heisenberg exchange coupling $J_1$ is calculated to be 3.184 meV for dn-MnIn$_2$Te$_4$ ML. The positive value of $J_1$ indicates that a ferromagnetic (FM) coupling in this system. The second NN exchange coefficient $J_2$, on the other hand, favors antiferromagnetic (AFM) ordering. It is remarkable that dn-MnIn$_2$Te$_4$ monolayer has a strong DMI and a significant $|d_1/J_1|$ ratio of 0.228, suggesting a high potential for generating skyrmionic phases. For up-MnIn$_2$Te$_4$, however, $J_1$ and $J_2$ are determined to be -6.312 and 0.184 meV, which indicate AFM coupling. Nevertheless, the value of $d_1$ for up-MnIn$_2$Te$_4$ ML is found to be only 0.224 meV, which yields $|d_1/J_1|$ = 0.035 and is too small to create topological spin textures. Magnetic anisotropy is also a crucial factor that influences the formation of magnetic topological states. In the case of dn-MnIn$_2$Te$_4$ ML, the anisotropic exchange interaction parameter λ is estimated to be 0.217 meV, indicating the out-of-plane magnetic anisotropy. The SIA parameter K is calculated to be -0.048 meV, which favors in-plane magnetic anisotropy. **It is apparent that λ plays a dominant role in estimating the out-of-plane magnetic anisotropy for dn-MnIn$_2$Te$_4$ ML.** Meanwhile, up-MnIn$_2$Te$_4$ ML exhibits λ = 0.301 meV and K = 0.893 meV, suggesting a rather stronger out-of-plane magnetic anisotropy. **If we look deeper into the structures of two types of MnIn$_2$Te$_4$ ML, we have: for dn-MnIn$_2$Te$_4$ ML, the bond length $l_1$ and $l_2$ is 2.679 and 2.736 Å, respectively, while or up-MnIn$_2$Te$_4$ ML, $l_1$ and $l_2$ are 2.649 and 2.843 Å, respectively.** The bond length $l_2$ can be regarded as the interlayer distance between MnTe layer and In$_2$Te$_3$ layer, where a smaller interlayer distance implies stronger interlayer interactions. To illustrate this, we carry out the Bader charge analysis [54]. Our results indicate that for dn-MnIn$_2$Te$_4$ ML, MnTe layer transfers 0.124 electrons to In$_2$Te$_3$, while for up-MnIn$_2$Te$_4$ ML, MnTe loses 0.077 electrons to In$_2$Te$_3$. This is also clearly reflected in the planar-average charge-density difference plotted in Figure 1 (f) and (g). Therefore, the distinct magnetic properties of dn- and up-MnIn$_2$Te$_4$ ML may arise from the difference in charge transfer between the MnTe and In$_2$Te$_3$ layers. From the band structures depicted in Figure 1 (d) and (e), both of the MnIn$_2$Te$_4$ ML exhibit metallic behavior. Thus, dn- and up-MnIn$_2$Te$_4$ ML is FM and AFM metal, respectively.

Once we obtained all the parameters in the spin Hamiltonian of Eq. (1), parallel tempering Monte Carlo (PTMC) simulations are then performed over a 60 × 60 × 1 supercell to explore spin structures of two types of MnIn$_2$Te$_4$ ML (see Note 1 in the Supporting Information for details). The initial spin configurations are obtained via full relaxation of the paramagnetic (random) state. To ensure that the converged spin structures to be the ground state at 0 K temperature, the conjugate gradient (CG) method is further applied to relax the spin configurations obtained through PTMC simulations.

As expected, due to the relatively small DMI of up-MnIn$_2$Te$_4$ ML, its ground state is determined to be out-of-plane 120° AFM, as shown in Figure S5. Surprisingly, despite having a substantial DMI, the ground state of dn-MnIn$_2$Te$_4$ ML is found to be out-of-plane FM. Previous studies have suggested that, in the case of FM coupled system, despite the presence of a significant DMI, the magnetic moments tend to align parallel to each other when a large total MAE (0.60 meV for dn-MnIn$_2$Te$_4$) exists [35]. Therefore, it is possible to obtain topological spin textures by tuning the balance between MAE and DMI.

To this end, within the rigid band model [55], we calculate the MAE against the shift of the Fermi level ($E_F$). As illustrated in Figure 2(a), for up-MnIn$_2$Te$_4$ ML, the MAE remains large in the out-of-plane direction and changes slowly as the $E_F$ is shifted upwards or downwards in a range of ± 0.2 eV. In contrast, for dn-MnIn$_2$Te$_4$ ML, the MAE changes significantly with respect to the $E_F$. By shifting the $E_F$ downwards by about 0.04 meV, the MAE is reduced to nearly 0 meV. As the $E_F$ continues to move downwards, the MAE could be even switched from out-of-plane to in-plane. Hence, it will be beneficial to manipulate the MAE of dn-MnIn$_2$Te$_4$ ML. One possibility for that may be through applying an external electric field, as it has been extensively demonstrated as an effective means of manipulating the MAE [55]. **A moderate external electric field is therefore applied perpendicular to the 2D plane of the dn-MnIn$_2$Te$_4$ ML to simulate the electric field effect.** As illustrated in Figure 2(b), the MAE is considerably diminished when an external electric field is applied along the -Z direction. The MAE value approaches zero when the magnitude of the electric field is 0.3 V/Å. Conversely, the MAE increases as the electric field strength directed along the +Z direction is enhanced (see Note 3 in the Supporting Information for details). The significant impact of the electric field on the MAE of dn-MnIn$_2$Te$_4$ ML can be attributed to the charge transfer occurring between the MnTe and In$_2$Te$_3$ layers, as is clearly evident from the results presented. Specifically, a greater transfer of charge from MnTe layer to In$_2$Te$_3$ layer results in a smaller MAE, while a lesser transfer of charge leads to a larger MAE. Moreover, the isotropic Heisenberg exchange coupling $J_1$ and $J_2$ exhibit a gradual increase as the electric field changes from 0 to 0.3 V/Å, while the effect of the electric field on the DMI is negligible, as depicted in Figure 2(c).

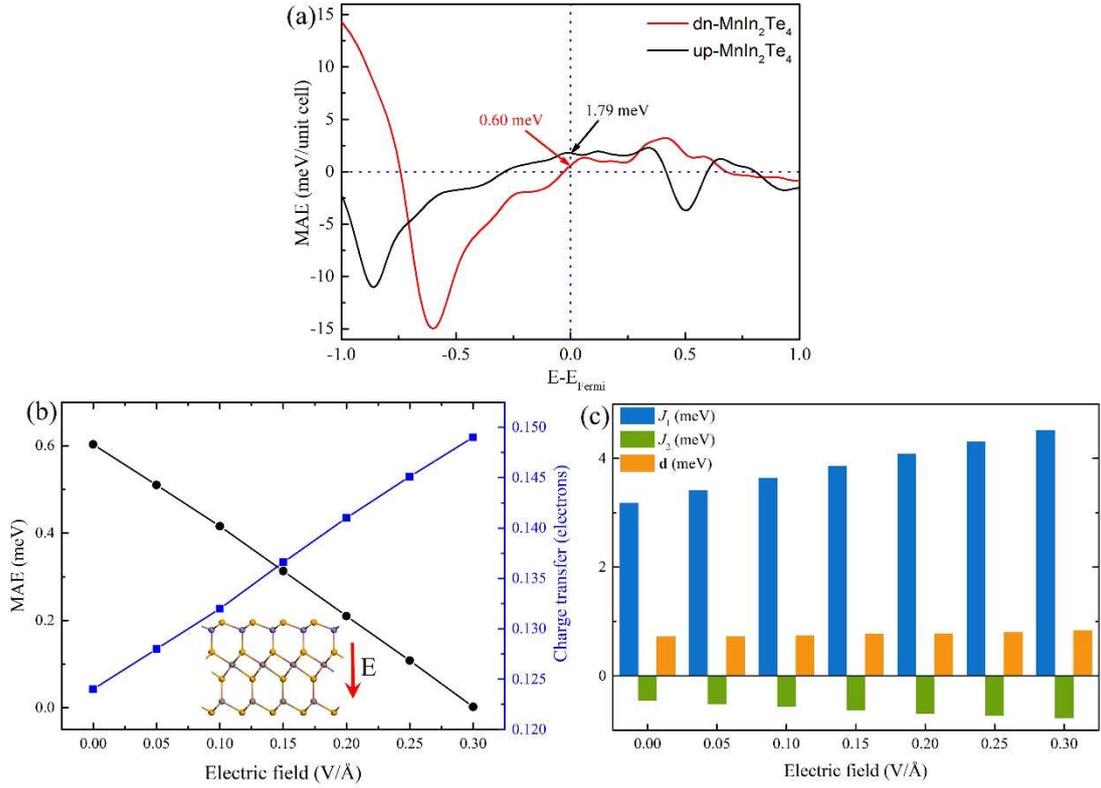

Figure 2. (a) MAE change as a function of the shift of Fermi energy for dn-MnIn$_2$Te$_4$ ML. The vertical dashed lines mark the real Fermi energy. (b) MAE of dn-MnIn$_2$Te$_4$ ML under an electric field from 0 to 0.3 V/Å. The insert indicates the direction of external electric field applied. (c) The Heisenberg exchange coupling $J_1$, $J_2$, and DMI as a function of electric field from 0 to 0.3 V/Å.

**The predicted electric control of MAE in the dn-MnIn$_2$Te$_4$ ML indicates a possibility of the electric field manipulation of magnetic skyrmions in this system.** To gain further insights into this effect, after calculating the magnetic parameters under different electric fields using DFT, the ground states of dn-MnIn$_2$Te$_4$ ML at 0 K without external magnetic field are first determined by employing the same strategy as used for the zero electric field case. We find that when the electric field is below 0.08 V/Å, the ground state of dn-MnIn$_2$Te$_4$ ML is determined to be out-of-plane ferromagnetism (FMz), of which the energy is set to be zero as the reference energy level. At an electric field of 0.08 V/Å, the ground state is identified as labyrinth domain, with an energy of -0.0022 meV per spin, as shown in Figure 3 (a). Under a higher electric field, for instance, at 0.12 V/Å, it can be clearly observed that the ground state of dn-MnIn$_2$Te$_4$ ML remains to be labyrinth domain, with the energy of -0.0341 meV per spin and a reduced domain width, as illustrated in Figure 3 (b).

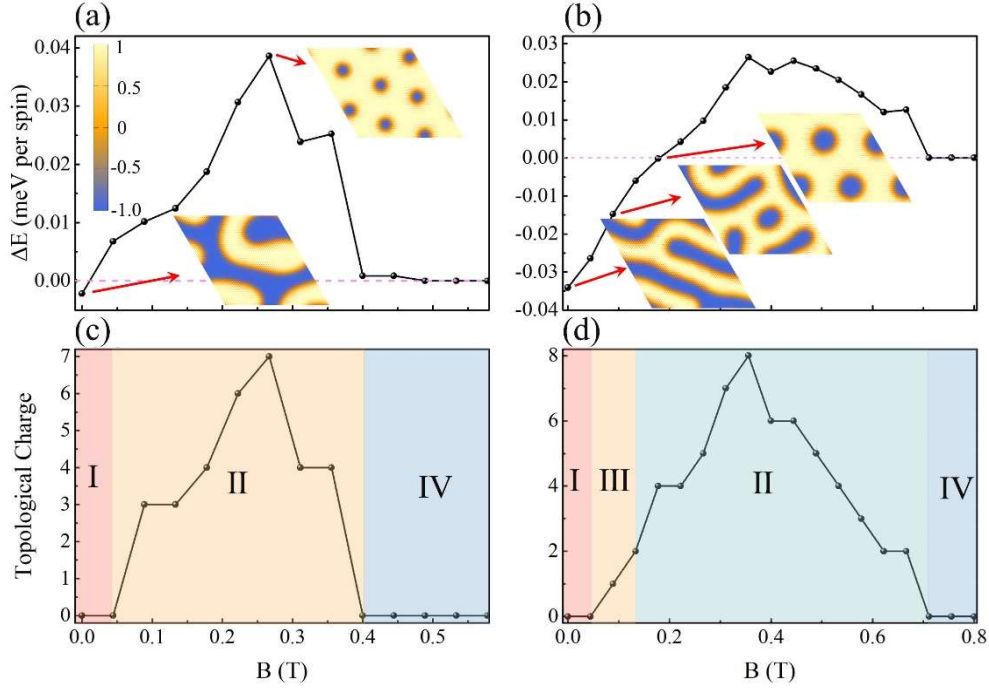

Figure 3. The energy difference ΔE of E(relax) and E(FMz) as a function of B at 0 K for electric field is (a) 0.08 and (b) 0.12 V/Å, respectively. Here E(relax) is the energy obtained from CG method. The inserts present the spin textures at different B. The topological charge |Q| as a function of B at 0 K for electric field is (c) 0.08 and (d) 0.12 V/Å, respectively. The phases found are as follows: labyrinth domains (I), discrete skyrmions/skyrmions lattice (II), labyrinths and skyrmion mixed phase (III), and saturated ferromagnetic state (IV), respectively.

We shall now investigate the impact of magnetic field on the topological spin structures of dn-$MnIn_2Te_4$ ML under different electric fields at 0 K. To this end, we have extracted the field-dependent values of topological charge Q per supercell from our simulations. Along with the real-space images presented in the inserts of Figure 3 (a) and (b), the values of Q can aid us in identifying the distinct phases. For instance, Q = 0 can correspond to stripes, labyrinths, or FM state, while nonzero values of Q signify the emergence of skyrmions. The external magnetic field is directed outward from the $In_2Se_3$ to MnTe layer. For the case of electric field is 0.08 V/Å, a phase denoted as phase II, characterized by discrete skyrmions, begins to appear at about 0.089 T, as indicated in Figure 3(c). It should be noted that this phase is metastable, and possesses a higher energy compared to the corresponding FM state. Above 0.4 T, the metastable skyrmion state disappears and the system transitions to a uniform FM state, which is referred to as phase IV. Interestingly, under electric field of 0.12 V/Å, we find that applying a magnetic field of 0.089 T leads to the appearance of a new phase III with lower energy compared to the FM state, where skyrmion is embedded within labyrinth domain. Upon increasing the magnetic field to 0.178 T, the labyrinth domains are totally wiped out and an ordered skyrmion lattice state emerges with a marginally lower energy (-0.0002 meV per spin) compared to the FM state. Skyrmions, on the other hand, appear as metastable state within the magnetic field range of 0.222 - 0.711 T. When the magnetic field exceeds 0.711 T, the system becomes FM state.

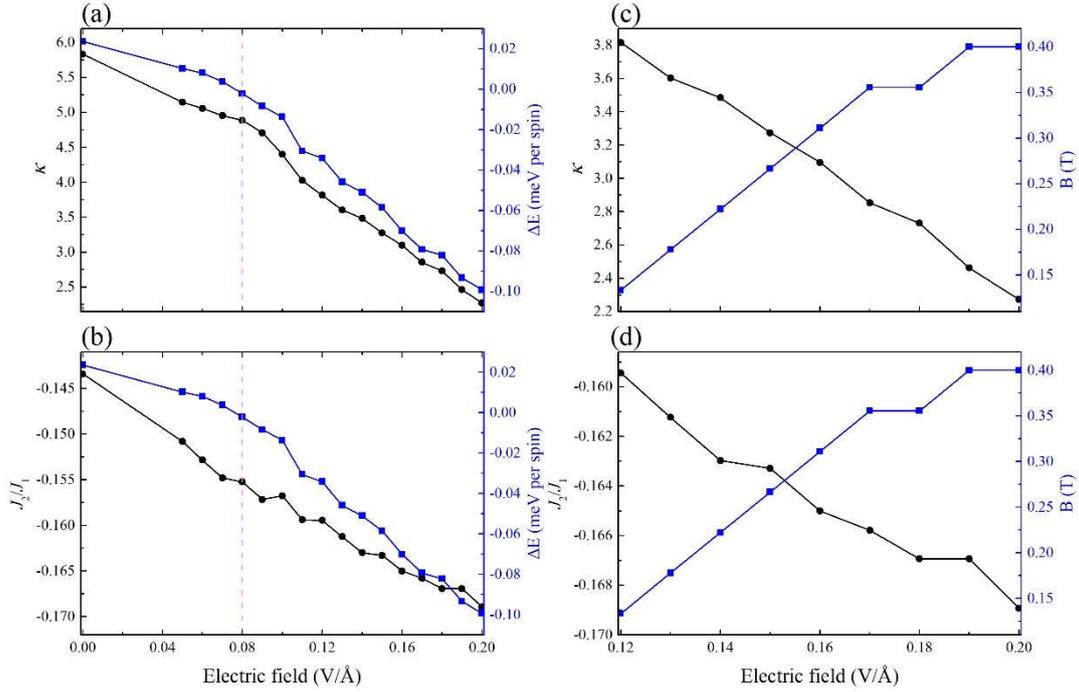

Figure 4. (a) κ and ΔE (right axes), and (b) $J_2/J_1$ ratio and ΔE (right axes) as a function of electric field from 0 to 0.3 V/Å, respectively. (c) κ and the critical magnetic field B that can form skyrmions, and (d) $J_2/J_1$ ratio and the critical magnetic field B as a function of electric field from 0 to 0.3 V/Å. The definition of ΔE is the same as Figure 3.

To deeply understand the electric-field-tunable skyrmionics physics in dn-MnIn$_2$Te$_4$ ML, we investigate the relationships between topological spin structures and magnetic parameters. Therefore, we first introduce the dimensionless parameter $\kappa = (\frac{4}{\pi})^2 \frac{J_1 * MAE}{D^2}$, that can help us identify magnetic ground state [56]. Specifically, when 0 < κ < 1, the magnetic ground state of the system exhibits spin spiral, whereas for κ > 1, the system tends to exhibit a collinear magnetic structure. The frustration arising from the competition between the second NN exchange coupling $J_2$, which favors AFM for all electric field values, and the FM $J_1$, as measured by the $J_2/J_1$ ratio, is another key factor that affects the magnetic structure of our system. We identify critical values of κ = 4.89 and $J_2/J_1$ = -1.56, below which the system's ground state tends to be spin spiral state, as illustrated in Figure 4(a) and (b). This critical κ value is close to that observed in the metal multilayers [16, 57]. Despite the deviation of κ from 1, the ground state remains spin spiral state due to the existence of frustration. We also consider the case without frustration and find that the ground state maintains spin spiral only when κ = 2.27 (see Note 7 in the Supporting Information for details). Our results are consistent with the previous reports that frustration can further stabilize the spiral/skyrmion state [32, 58-60]. From a practical application standpoint, it is found that a stronger magnetic field B is necessary to form pure skyrmions when κ and $J_2/J_1$ are smaller, as demonstrated in Figure 4 (c) and (d). Thus, by applying an appropriate external magnetic field, the electrically controllable skyrmionics can be realized.

To summarize, we propose that dn-$MnIn_2Te_4$ ML is appealing 2D magnets with skyrmionics physics based on first-principles calculations and Monte Carlo simulations. Upon applying an electric field from 0 to 0.3 V/Å, except the DMI that shows negligible response to the electric field, the MAE can be significantly decreased and isotropic Heisenberg exchange coupling $J_1$ and $J_2$ exhibit a gradual increase, which can be attributed to the charge transfer occurring between the MnTe and $In_2Te_3$ layers. Thus, by applying an electric field, the creation and annihilation of skyrmion state can be realized, giving rise to the exotic electrically controllable skyrmionics. Such phenomenon is revealed to be closely related to the physical quantities of κ and $J_2/J_1$ ratio, which can describe the required conditions for such physics. Our work demonstrates a possibility of the nonvolatile control of magnetic skyrmions in 2D magnets by applying electric field, which is promising for the potential device application of skyrmion systems.

**Associated Content**

Supporting Information

The Supporting Information is available free of charge at ***

Details of method, skyrmion number Q, Phonon spectra, AIMD simulation results,

effect of electric field along +Z direction, 120° AFM spin texture of dn-MnIn$_2$Te$_4$ ML, effect of structure relaxation under electric field, spin texture of electric field equals 0.2 V/Å, the case of only including the NN $J_1$, effect of Hubbard U values.

**Author Information**

**Acknowledgments**

This work is supported by the Natural Science Basic Research plan in Shaanxi Province of China (2022JQ-008, 2023-JC-YB-021), Open Foundation of Key Laboratory of Computational Physical Sciences (Ministry of Education), and Open Foundation of State Key Laboratory of Surface Physics and Department of Physics (KF2022_14). L.B. thanks the Vannevar Bush Faculty Fellowship Grant No. N00014-20-1-2834 from the Department of Defense. Calculations were performed at the High-Performance Computing Center of Baoji University of Arts and Sciences.